\documentclass[aps,preprint]{revtex4}
\usepackage{amsfonts}
\usepackage{amsmath}
\usepackage{amssymb}
\usepackage{graphicx}
\usepackage[colorlinks,hyperindex]{hyperref}

\hypersetup{
colorlinks,
citecolor=blue,
linkcolor=black,
urlcolor=black,
}

\begin{document}

\title[Two-dimensional PDM particles]{A complete set of eigenstates for
position-dependent massive particles in a Morse-like scenario}
\author{R. A. C. Correa\thanks{%
Electronic Address: rafael.couceiro@ufabc.edu.br}}
\affiliation{Universidade Estadual Paulista-UNESP, Campus de Guaratinguet\'{a},
12516-410, Guaratinguet\'{a}, SP, Brazil}
\author{A. de Souza Dutra\thanks{%
Electronic Address: dutra@feg.unesp.br}}
\affiliation{Universidade Estadual Paulista-UNESP, Campus de Guaratinguet\'{a},
12516-410, Guaratinguet\'{a}, SP, Brazil}
\author{J. A. de Oliveira\thanks{%
Electronic Address: julianoantonio@sjbv.unesp.br}}
\affiliation{Universidade Estadual Paulista-UNESP, Campus de S\~{a}o Jo\~{a}o da Boa
Vista, 13874-149, S\~{a}o Jo\~{a}o da Boa Vista, SP, Brazil.}
\author{M. G. Garcia\thanks{%
Electronic Address: marcelogarcia82@hotmail.com }}
\affiliation{Universidade Estadual Paulista (UNESP), Campus de Guaratinguet\'{a},
12516-410, Guaratinguet\'{a}, SP, Brazil}
\keywords{Morse, Two-Dimensional, Position-Dependent}
\pacs{03.65.Ge, 03.65.-w}

\begin{abstract}
In this work we analyze a system consisting in two-dimensional
position-dependent massive particles in the presence of a Morse-like
potential in two spatial dimensions. We obtain the exact wavefunctions and
energies for a complete set of eigenstates for a given dependence of the
mass with the spatial variables. Furthermore, we argue that this scenario
can be play an important role to construct more realistic ones by using
their solution in perturbative approaches.
\end{abstract}

\date{October 13, 2016}
\maketitle


\section{Introduction}

Some years ago, the materials science took an important step forward by
obtaining the fabrication of small conducting devices known as quantum dots
(QDs) \cite{Alhassid}. In those devices, it is possible to confine several
thousand electrons in a small region whose linear size is about $0.1-1\mu m$ 
\cite{rmp1}. The fundamental characteristic of QDs is that they are
typically formed by a two-dimensional electron gas, where, by applying an
electrostatic potential, the electrons are confined to a small region, which
is called \textquotedblleft dot\textquotedblright , in the interface region
of a semiconductor. A very important advantage of QDs is that their
transport properties are readily measured, allowing an experimental control.
Moreover, the effects of time-reversal symmetry breaking can be easily
measured by applying a magnetic field \cite{prl0}. Nowadays, a variety of
theoretical and experimental research about small conducting devices, such
as QDs, has been the focus of many scientists and engineers attention \cite%
{prb1,prb2,pre1,prl1}. From a phenomenological viewpoint, QDs are very small
structures, where the laws of quantum mechanics (QM) are the most important
ingredients to describe their properties. Thus, as\ a natural consequence of
practical applicability of the theoretical framework of QM, a great interest
arises for exact solutions of two-dimensional confined systems, which can be
fundamental to explore the physics in small conducting devices, such as the
QDs. In the light of these facts, it was shown in Ref. \cite{Lozada-Dong-Yu}
that it is possible to find exact solutions of the two-dimensional Schr\"{o}%
dinger equation with the position-dependent mass (PDM) for the square well
potential in the semiconductor quantum dots (SQDs) system. Another important
work in this context, it was presented by Schmidt, Azeredo, and Gusso \cite%
{Schmidt-Azeredo-Gusso}, where the authors have studied both the problems of
quantum wave packet revivals on two-dimensional infinite circular quantum
wells (CQWs) and circular quantum dots (CQDs) with PDM, showing the results
for the eigenfunctions, eigenenergies and the revival time for spatially
localized electronic Gaussian wave packets. At this point, it is important
to highlight that the importance in adding a PDM is due to the fact that the
system will take into account the spatial variation of the semiconductor 
\cite{BenDaniel-Duke, Gora-Williams, Bastard, Ross, Zhu-Kroemer, Li-Kuhn,
Cavalcante-Filho-Almeida-Freire, Dutra-Oliveira}. However, as consequence of
inclusion a PDM, the system becomes ambiguous at the quantum level, and the
ordering ambiguity problem (OAP) is one of the long standing unsolved
questions in quantum mechanics. As we know the OAP has attracted the
attention of some of the founders of the quantum mechanics, namely, Born,
Jordan, Weyl, Dirac and von Newmann worked on this problem, as can be
verified from the review by Shewell \cite{Shewell}. This is viewed as a deep
problem in QM, which has advanced very few along the last decades. But all
is not lost, it was shown that the ordering ambiguous problem has a very
special importance for the modeling of some experimental situations like
electrons in perturbed periodic lattices \cite{Slater}, impurities states
and cyclotron resonance in semiconductors \cite{Luttinger}, the structure of
electronic excitation levels in insulating crystals \cite{Wannier},\ the
dependence of nuclear forces on the relative velocity of the two nucleons 
\cite{rojo, razavy}, and more recently the study of semiconductor
heterostructures \cite{Bastard,weisbuch,ref1,ref2,ref3}. Moreover, some time
ago, it was discussed in the literature the exact solvability of some
classes of one-dimensional Hamiltonians, where the potentials has a PDM,
with ordering ambiguity \cite{pla2000}, after that, a large number of works
regarding one-dimensional Hamiltonians with ordering ambiguity has emerged
in the scientific community along the last few years \cite%
{m1,m2,m3,m4,m5,m6,jpa2006}. Another interesting research line regards to
the supersymmetry approach to one-dimensional quantum systems with
spatially-dependent mass, by including their ordering ambiguities dependence 
\cite{dutra2, schimidt1, quesne, sever, halberg, carinena, ganguly, roy,
mustafa, tanaka, koca, znojil, cg1, Cg2, a0, a1, a3, a4, a5, a6, a7, a8, a9,
npb, epl05}. On the other hand, as far as we know, some physical systems
like ones where a magnetic field is present \cite{pla87, pra89, pla91,
abdalla2007}, lead naturally to the necessity of a two-dimensional analysis.
In the face of this situation, it was presented in Ref.\cite{dutra-juliano}
a general approach for the problem of a particle with PDM interacting with a
two-dimensional potential well with finite depth, where the ordering
ambiguity was taken in account. In that work, it was shown that the
considered system retain an infinite set of quantum states, which usually do
not happens in the case of the constant mass systems. Furthermore, it was
verified also that the SU(2) coherent state corresponds to a stationary
state. Also recently, numerous other theoretical studies have been conducted
on two-dimensional position-dependent mass Schr\"{o}dinger equation (PDMSE).
These include the two-dimensional quantum rotor with two effective masses 
\cite{cc1}, kinetic operator in cylindrical coordinates \cite{cc2}, exact
solutions for the PDMSE in an annular billiard with impenetrable walls \cite%
{cc3}, and a particle with spin 1/2 moving in a plane \cite{cc4}.

Here, we will address the position-dependent mass (PDM) type of the systems
in two spatial dimensions (2D) by using Cartesian coordinates. We will
introduce a very interesting system where, as we are going to see below in
the manuscript, the relation between the quantum numbers introduced along
the procedure of resolving the equations of the system and the energy
eigenstates organization is somewhat remarkable.

This paper is organized as follows. In Section \textcolor{red}{II}, we
review the effective Schr\"{o}dinger equation in two-dimensional Cartesian
coordinates. In Section \textcolor{red}{III}, we introduce the
Position-dependent massive particle with Morse-like terms and its exact
solutions. In Section \textcolor{red}{IV}, we present our conclusions and
directions for future work.

\section{Effective Schroedinger equation in two-dimensional Cartesian
coordinates: A brief review}

In this section, we will recapitulate the results presented some years ago
in Ref. \cite{Dutra-Oliveira}. Let us start with the ordering defined by von
Roos \cite{Ross, pla2000} for the Hamiltonian operator, which in
one-dimensional space is written in the following form 
\begin{equation}
\hat{H}=\frac{1}{4}\left( M^{\alpha}\hat{p}M^{\beta}\hat{p}M^{\gamma
}+M^{\gamma}\hat{p}M^{\beta}\hat{p}M^{\alpha}\right) +V(x),  \label{eq2a}
\end{equation}
where $\hat{p}$ is the momentum operator and $M=M(x)$ is the
position-dependent effective mass. Moreover, $\alpha,\beta$ and $\gamma$ are
arbitrary ordering parameters which must to obey the relation 
\begin{equation}
\alpha+\beta+\gamma=-1.  \label{eq2b}
\end{equation}

At this point, it is important to highlight that the above relation is
necessary to get the correct classical limit.

Applying the canonical commutation relations, we have

\begin{equation}
M^{\gamma}\hat{p}M^{\beta}\hat{p}M^{\alpha}=\frac{\hat{p}^{2}}{M}-i\hbar
(\beta+2\alpha)\frac{M^{\prime}}{M^{2}}\hat{p}-\hbar^{2}\alpha(\beta
+\alpha-1)\frac{(M^{\prime})^{2}}{M^{3}}-\hbar^{2}\alpha\frac{%
M^{\prime\prime }}{M^{2}}.  \label{eq2q}
\end{equation}

Through the relation (\ref{eq2b}), the effective Hamiltonian operator \cite%
{pla2000} is given by

\begin{equation}
H=\frac{1}{2\,M}\hat{p}^{2}+\frac{i\,\hbar}{2}\frac{M^{\prime}}{M^{2}}\hat {p%
}+U(\alpha,\gamma,x)+V(x),  \label{eq2ss}
\end{equation}

\noindent where the effective potential $U(\alpha,\gamma,x)$ is written as

\begin{equation}
U(\alpha,\gamma,x)=-\frac{\hbar^{2}}{4M^{3}}\left[ (\alpha+\gamma)M\left( 
\frac{\partial^{2}M}{\partial x^{2}}\right) -2(\alpha+\gamma+\alpha
\gamma)\left( \frac{\partial M}{\partial x}\right) ^{2}\right] .  \label{1}
\end{equation}

Therefore, we can now write the effective Schroedinger equation in the form

\begin{equation}
-\frac{\hbar^{2}}{2\,M(x)}\frac{d^{2}\psi}{dx^{2}}+\frac{\hbar^{2}}{2}\left[ 
\frac{dM/dx}{M^{2}}\right] \frac{d\psi}{dx}+[V(x)+U(\alpha,\gamma
,x)-E]\psi=0.  \label{eq2ww}
\end{equation}

In the case of a set of two-dimensional Cartesian coordinates, where $%
M=M(x,y)$, the effective Hamiltonian operator

\begin{equation}
H=\frac{1}{2\,M(x,y)}(\hat{p_{x}}^{2}+\hat{p_{y}}^{2})+\frac{i\,\hbar}{2}%
\left( \frac{\frac{\partial M}{\partial x}\hat{p_{x}}+\frac{\partial M}{%
\partial y}\hat{p_{y}}}{M^{2}}\right) +U(\alpha.\gamma,x)+V(x,y),
\label{eq2s}
\end{equation}

\noindent where $U(\alpha,\gamma,x,y)$ is the effective potential. Now it
can be written in the form \cite{Dutra-Oliveira}

\begin{equation}
U(\alpha,\gamma,x,y)=-\frac{\hbar^{2}}{4M}\left\{ (\alpha+\gamma )\,\frac{%
M_{xx}+M_{yy}}{M}-2(\alpha+\gamma+\alpha\gamma)\left[ \left( \frac{M_{x}}{M}%
\right) ^{2}+\left( \frac{M_{y}}{M}\right) ^{2}\right] \right\} ,
\label{eq2t}
\end{equation}

\noindent with $M_{x}\equiv\partial M/\partial x$ and $M_{y}\equiv\partial
M/\partial y$. Therefore, we have

\begin{equation}
H=\frac{1}{2\,M}\overrightarrow{p}^{2}+\frac{i\hbar}{2}\,\frac{1}{M^{2}}\,%
\overrightarrow{\nabla}M.\overrightarrow{p}+U(\alpha,\gamma,x,y)+V(x,y),
\label{eq2u}
\end{equation}
\noindent where, in this case 
\begin{equation}
U(\alpha,\gamma,x,y)\equiv-\frac{\hbar^{2}}{4\,M}\left[ (\alpha +\gamma)\,%
\frac{{\nabla}^{2}M}{M}-2(\alpha+\gamma+\alpha\gamma)\left( \frac{%
\overrightarrow{\nabla}M}{M}\right) ^{2}\right] .  \label{eq2v}
\end{equation}

In the next step we can use a typical Schroedinger equation 
\begin{equation}
-\frac{\hbar^{2}}{2\,M(x,y)}\,\nabla^{2}\,\chi+V_{eff}(x,y)\,\chi=E\,\chi,
\label{eq2w}
\end{equation}
and if $\chi(x,y)=e^{\sigma(x,y)}\,\psi(x,y)$ is the solution of it, the
equation above can be rewritten as follows

\begin{align}
& -\frac{\hbar^{2}}{2\,M(x,y)}\nabla^{2}\psi-\frac{\hbar^{2}}{M(x,y)}\left[
\left( \overrightarrow{\nabla}\sigma\right) .\overrightarrow{\nabla}\psi%
\right] +  \notag  \label{eq2d1} \\
& +\left\{ V(x,y)-\frac{\hbar^{2}}{2\,M(x,y)}\left[ \nabla^{2}\sigma+\left( 
\overrightarrow{\nabla}\sigma\right) ^{2}\right] \right\} \psi=E\psi.
\end{align}

The above equation have a Hamiltonian operator defined by

\begin{equation}  \label{eq2e1}
H = \frac{1}{2\,M(x,y)} \overrightarrow{p}^{2} -\frac{\hbar^{2}}{M(x,y)} \, 
\frac{i}{\hbar} (\overrightarrow{\nabla} \sigma) . \overrightarrow{p} + V - 
\frac{\hbar^{2}}{2\,M(x,y)}\left[ \nabla^{2} \sigma+ (\overrightarrow{\nabla}
\sigma)^{2}\right] .
\end{equation}

Note that we can choose 
\begin{equation}
-\frac{\hbar^{2}}{M}\,\frac{i}{\hbar}\,\overrightarrow{\nabla}\sigma .%
\overrightarrow{p}=\frac{i\hbar}{2}\,\frac{1}{M^{2}}\,\overrightarrow{\nabla 
}M.\overrightarrow{p},  \label{eq2f1}
\end{equation}

\noindent such that

\begin{equation}  \label{eq2g1}
\frac{\overrightarrow{\nabla} M}{M} = -2 \,\overrightarrow{\nabla } \sigma.
\end{equation}

Thus, we have%
\begin{equation}
\sigma=ln(M^{-\frac{1}{2}}).
\end{equation}

Now, we may rewrite the equation (\ref{eq2e1}) as

\begin{equation}
H=\frac{1}{2\,M}\overrightarrow{p}^{2}+\frac{i\hbar}{2\,M}\frac {%
\overrightarrow{\nabla}\,M}{M}.\overrightarrow{p}+\left\{ V-\frac{\hbar^{2}}{%
4\,M}\left[ \frac{3}{2}\left( \frac{\overrightarrow{\nabla}M}{M}\right) ^{2}-%
\frac{\nabla^{2}M}{M}\right] \right\} .  \label{e2}
\end{equation}

On the other hand, the wavefunction is re-scaled as

\begin{equation}  \label{eq2m1}
\psi= M^{\frac{1}{2}} \, \chi.
\end{equation}

From these results, we see that \cite{Dutra-Oliveira}

\begin{equation}
V_{eff}(x,y)=V(x,y)+\frac{\hbar^{2}}{4M}\left[ 2\left( \alpha+\gamma
+\alpha\gamma+\frac{3}{4}\right) \left( \frac{\overrightarrow{\nabla}M}{M}%
\right) ^{2}-\left( \alpha+\gamma+1\right) \,\frac{\nabla^{2}M}{M}\right] .
\label{eq2r1}
\end{equation}

Finally, we must comment that for an equivalent system with constant mass,
the equation (\ref{eq2w}) can be written as

\begin{equation}  \label{eq2s1}
-\frac{\hbar^{2}}{2} \nabla^{2} \chi+ U_{eff} \, \chi= \xi\chi,
\end{equation}

\noindent with $\xi$ constant and

\begin{align}
U_{eff}-\xi & =M(x,y)\,V(x,y)+\frac{\hbar^{2}}{4}\left[ 2\left(
\alpha+\gamma+\alpha\gamma+\frac{3}{4}\right) \left( \frac {\overrightarrow{%
\nabla}M}{M}\right) ^{2}+\right.  \notag \\
& \left. -\left( \alpha+\gamma+1\right) \,\frac{\nabla^{2}M}{M}\right]
-E\,M(x,y),  \label{mass}
\end{align}

Through the above result, it was studied in \cite{Dutra-Oliveira} the
problem of a particle with a position-dependent mass interacting with
two-dimensional potential well with finite depth, as well as under the
influence of a uniform magnetic field. There, it was discovered that the
system retains an infinite set of quantum states. In the next section, we
explore the problem where the PDM is Morse-like.

\section{Position-dependent massive particle with Morse-like terms}

An important problem in quantum mechanics is that one related to the
vibrations of diatomic molecules, and the case of vibrations of a two-atomic
molecule are well described by the Morse potential. On the other hand, there
is a growing number of applications of quantum wells and quantum dots. In
fact, those systems present a small spatial region capable to confine
quantum particles. As one can see in Figure 1, the bidimensional Morse-like
potential can simulate such kind of physical situation and it has the
advantage, as we will see below, of being exactly solved. Furthermore,
having the exact solutions in hands, one can use them in order to describe
more realistic problems by using approximation techniques which make use of
those exact solutions. Therefore, with this motivation in our mind, in this
section, let us present an example which can be exactly solved. Thus, we
will consider that

\begin{equation}
M(x,y)=M_{0}\left[ 1+g_{1}e^{-\alpha_{1}x}+g_{3}e^{-\alpha_{2}y}+g_{2}e^{-2%
\alpha_{1}x}+g_{4}e^{-2\alpha_{2}y}\right] .  \label{eq2v1}
\end{equation}

Note that the spatial dependence of the mass is similar to that of a Morse
potential in two dimensions. Thus, plugging this mass in the formula given
by (\ref{mass}), we obtain

\begin{align}
& \left. U_{eff}-\xi=M_{0}\left[ 1+g_{1}e^{-\alpha_{1}x}+g_{3}e^{-%
\alpha_{2}y}+g_{2}e^{-2\alpha_{1}x}+g_{4}e^{-2\alpha_{2}y}\right]
\,V(x,y)\right.  \notag \\
&  \notag \\
& \left. -EM_{0}\left[ 1+g_{1}e^{-\alpha_{1}x}+g_{3}e^{-%
\alpha_{2}y}+g_{2}e^{-2\alpha_{1}x}+g_{4}e^{-2\alpha_{2}y}\right] \right. 
\notag \\
&  \notag \\
& \left. +\frac{\hbar^{2}}{4}\left\{ 2\left( \alpha+\gamma+\alpha \gamma+%
\frac{3}{4}\right) \,\,\left[ \frac{\alpha_{1}^{2}g_{1}^{2}e^{-2%
\alpha_{1}x}+\alpha_{2}^{2}g_{3}e^{-2\alpha_{2}y}+4\alpha_{1}^{2}g_{2}e^{-4%
\alpha_{1}x}+4\alpha_{2}^{2}g_{4}e^{-4\alpha_{2}y}}{(1+g_{1}e^{-%
\alpha_{1}x}+g_{3}e^{-\alpha_{2}y}+g_{2}e^{-2\alpha_{1}x}+g_{4}e^{-2%
\alpha_{2}y})^{2}}\right] \right. \right.  \notag \\
&  \notag \\
& \left. \left. -(\alpha+\gamma+1)\left[ \frac{\alpha_{1}^{2}g_{1}e^{-2%
\alpha_{1}x}+\alpha_{2}^{2}g_{3}e^{-2\alpha_{2}y}+4\alpha_{1}^{2}g_{2}e^{-4%
\alpha_{1}x}+4\alpha_{2}^{2}g_{4}e^{-4\alpha_{2}y}}{1+g_{1}e^{-%
\alpha_{1}x}+g_{3}e^{-\alpha_{2}y}+g_{2}e^{-2\alpha_{1}x}+g_{4}e^{-2%
\alpha_{2}y}}\right] \right\} .\right.
\end{align}

In order to work with a exactly solvable model we can assume the following
ordering

\begin{equation}
\alpha+\gamma+1=0,\text{ }\alpha+\gamma+\alpha\gamma+\frac{3}{4}=0,
\label{eq2d2}
\end{equation}

\noindent whose solution is given by

\begin{equation}
\alpha=-\frac{1}{2},\text{ }\gamma=-\frac{1}{2},\text{ }\beta=0.
\label{eq2h2}
\end{equation}

In this way, we then obtain

\begin{equation}
\frac{\overrightarrow{p}^{2}}{2\,M}=\frac{1}{2}\frac{1}{\sqrt{M}}(%
\overrightarrow{p})^{2}\frac{1}{\sqrt{M}}.  \label{eq2k2}
\end{equation}

Consequently, in this ordering, the effective potential is written as

\begin{align}
U_{eff}-\xi & =M_{0}\left[ 1+g_{1}e^{-\alpha_{1}x}+g_{3}e^{-%
\alpha_{2}y}+g_{2}e^{-2\alpha_{1}x}+g_{4}e^{-2\alpha_{2}y}\right]
\,V(x,y)-EM_{0}  \notag \\
&  \notag \\
&
-EM_{0}(g_{1}e^{-\alpha_{1}x}+g_{3}e^{-\alpha_{2}y}+g_{2}e^{-2%
\alpha_{1}x}+g_{4}e^{-2\alpha_{2}y}).  \label{4}
\end{align}

As an example, we can choose a potential under which the particle with
position-dependent mass is moving. Then, here we will work with the
following potential%
\begin{equation}
V(x,y)=R+\frac{A+B_{1}e^{-\alpha_{1}x}+B_{3}e^{-\alpha_{2}y}+B_{2}e^{-2%
\alpha_{1}x}+B_{4}e^{-2\alpha_{2}y}}{M_{0}\left(
1+g_{1}e^{-\alpha_{1}x}+g_{3}e^{-\alpha_{2}y}+g_{2}e^{-2%
\alpha_{1}x}+g_{4}e^{-2\alpha_{2}y}\right) },  \label{eq2s2}
\end{equation}

\noindent where $R$ is a constant. In Figure 1, we plot a typical case where
this potential can confine particles. Therefore, through this potential, we
can obtain the effective potential

\begin{align}
U_{eff}-\xi & =A+M_{0}(R-E)+\left[ B_{1}+M_{0}(R-E)g_{1}\right]
e^{-\alpha_{1}x}+\left[ B_{2}+M_{0}(R-E)g_{2}\right] e^{-2\alpha_{1}x} 
\notag \\
&  \notag \\
& +\left[ B_{3}+M_{0}(R-E)g_{3}\right] e^{-\alpha_{2}y}+\left[
B_{4}+M_{0}(R-E)g_{4}\right] e^{-2\alpha_{2}y},
\end{align}

\noindent where we can easily see that $\xi=-A+M_{0}(E-R)$. So that

\begin{equation}
U_{eff}=\gamma_{1}e^{-\alpha_{1}x}+\gamma_{2}e^{-2\alpha_{1}x}+\gamma
_{3}e^{-\alpha_{2}y}+\gamma_{4}e^{-2\alpha_{2}y},
\end{equation}

\noindent with%
\begin{equation}
\gamma_{i}\equiv B_{i}+M_{0}(R-E)g_{i},\text{ }i=1,2,3,4.,
\end{equation}

Now, the Schroedinger equation (\ref{eq2s1}) takes the form%
\begin{equation}
-\nabla^{2}\chi+\frac{2}{\hbar^{2}}(\gamma_{1}e^{-\alpha_{1}x}+\gamma
_{2}e^{-2\alpha_{1}x}+\gamma_{3}e^{-\alpha_{2}y}+\gamma_{4}e^{-2%
\alpha_{2}y})\chi=\varepsilon~\chi.  \label{5}
\end{equation}

\noindent where $\varepsilon\equiv2\xi/\hbar^{2}$.

In order to solve the above equation, we can use the usual procedure of
variable separation%
\begin{equation}
\chi(x,y)=X(x)Y(y).
\end{equation}

Thus, we get the equations for $X(x)$ and $Y(y)$ below%
\begin{align}
-\frac{d^{2}X(x)}{dx^{2}}+(\eta_{1}e^{-\alpha_{1}x}+\nu_{1}e^{-2%
\alpha_{1}x})X(x) & =\varepsilon_{m}X(x),  \label{6} \\
&  \notag \\
-\frac{d^{2}Y(y)}{dy^{2}}+(\eta_{2}e^{-\alpha_{2}y}+\nu_{2}e^{-2%
\alpha_{2}y})Y(y) & =\varepsilon_{n}Y(y).  \label{7}
\end{align}

\noindent where 
\begin{equation}
\eta_{1}\equiv\frac{2\gamma_{1}}{\hslash^{2}},\text{ }\nu_{1}\equiv \frac{%
2\gamma_{2}}{\hslash^{2}},~\eta_{2}\equiv\frac{2\gamma_{3}}{\hslash^{2}},%
\text{ }\nu_{2}\equiv\frac{2\gamma_{4}}{\hslash^{2}}.
\end{equation}

Furthermore, the energy spectrum is given by%
\begin{equation}
\varepsilon_{mn}=\varepsilon_{m}+\varepsilon_{n}.  \label{8}
\end{equation}

Let us now determine the solution of $X(x)$. Note that the equation (\ref{7}%
) have the same form of (\ref{6}), of course, written in terms of the
variable $y$. In this way, it is necessary to solve only (\ref{6}). Thus, we
define the variable $z$ and constants $\mu$ and $\lambda$ as

\begin{equation}
z:=\frac{2\sqrt{\left\vert \nu_{1}\right\vert }}{\alpha_{1}}e^{-\alpha_{1}x},%
\text{ }\mu:=\frac{\sqrt{\left\vert \varepsilon_{m}\right\vert }}{\alpha_{1}}%
,\text{ }\lambda:=-\frac{\eta_{1}}{2\alpha_{1}\sqrt{\left\vert
\nu_{1}\right\vert }}.
\end{equation}

\noindent with -$\infty<x<\infty$. In this case, bound states are possible
only for $\nu_{1}>0$ and $\eta_{1}<0$. Then, we have%
\begin{equation}
\varepsilon_{m}=-\frac{1}{4\nu_{1}}\left[ \left\vert \eta_{1}\right\vert
-\alpha_{1}\sqrt{\nu_{1}}(2m+1)\right] ^{2},\text{ with }m=0,1,2,3,...,m_{%
\max}.
\end{equation}

Furthermore, the function $X(x)$ is given by%
\begin{equation}
X_{m}(x)=\left( \frac{2\sqrt{\left\vert \nu_{1}\right\vert }}{\alpha_{1}}%
\right) ^{\mu}\exp\left[ -\left( \mu\alpha_{1}x+\frac{\sqrt{\left\vert
\nu_{1}\right\vert }}{\alpha_{1}}e^{-\alpha_{1}x}\right) \right] \mathcal{L}%
_{m}^{2\mu}\left( z\rightarrow\frac{2\sqrt{\left\vert \nu _{1}\right\vert }}{%
\alpha_{1}}e^{-\alpha_{1}x}\right) ,  \label{9}
\end{equation}

\noindent where $\mathcal{L}_{m}^{2\mu}(x)$ are the Laguerre polynomials.
Here, it is important to remark that the number of discrete levels is finite
and determined by the condition%
\begin{equation}
\left\vert \eta_{1}\right\vert >\alpha_{1}\sqrt{\nu_{1}}(2m_{\max}+1).
\end{equation}
This happens due to the fact that the potential goes asymptotically to zero
when $x\longrightarrow\infty$ and its minimum value is negative. So the
energy levels for bounded particles must be lower than zero, which leads to
the above constraint. On the other hand, defining%
\begin{equation*}
\bar{z}:=\frac{2\sqrt{\left\vert \nu_{1}\right\vert }}{\alpha_{1}}%
e^{-\alpha_{1}x},\text{ }\bar{\mu}:=\frac{\sqrt{\left\vert \varepsilon
_{n}\right\vert }}{\alpha_{2}},\text{ }\bar{\lambda}:=-\frac{\eta_{2}}{%
2\alpha_{2}\sqrt{\left\vert \nu_{2}\right\vert }},
\end{equation*}

\noindent and solving the equation (\ref{7}), we obtain

\begin{equation}
\varepsilon_{n}=-\frac{1}{4\nu_{2}}\left[ \left\vert \eta_{2}\right\vert
-\alpha_{2}\sqrt{\nu_{2}}(2n+1)\right] ^{2},\text{ with }%
n=0,1,2,3,...,n_{mx}.
\end{equation}

\noindent and

\begin{equation}
Y_{m}(y)=\left( \frac{2\sqrt{\left\vert \nu_{2}\right\vert }}{\alpha_{2}}%
\right) ^{\bar{\mu}}\exp\left[ -\left( \bar{\mu}\alpha_{2}y+\frac {\sqrt{%
\left\vert \nu_{2}\right\vert }}{\alpha_{2}}e^{-\alpha_{2}y}\right) \right] 
\mathcal{L}_{n}^{2\bar{\mu}}\left( \bar{z}\rightarrow\frac {2\sqrt{%
\left\vert \nu_{2}\right\vert }}{\alpha_{2}}e^{-\alpha_{2}y}\right) ,
\end{equation}

\noindent with the condition%
\begin{equation}
\left\vert \eta_{2}\right\vert >\alpha_{2}\sqrt{\nu_{2}}(2n_{\max}+1).
\end{equation}

Therefore, the total energy is written as%
\begin{equation}
\varepsilon_{mn}=-\frac{1}{4\nu_{1}\nu_{2}}\left\{ \nu_{2}\left[ \left\vert
\eta_{1}\right\vert -\alpha_{1}\sqrt{\eta_{1}}(2m+1)\right] ^{2}+\nu _{1}%
\left[ \left\vert \eta_{2}\right\vert -\alpha_{2}\sqrt{\eta_{2}}(2n+1)\right]
^{2}\right\} .
\end{equation}

Moreover, we write

\begin{align}
& \left. \chi_{mn}(x,y)=\left( \frac{2\sqrt{\left\vert \nu_{1}\right\vert }}{%
\alpha_{1}}\right) ^{\mu}\left( \frac{2\sqrt{\left\vert \nu _{2}\right\vert }%
}{\alpha_{2}}\right) ^{\bar{\mu}}\exp\left\{ -\left[ \left( \mu\alpha_{1}x+%
\frac{\sqrt{\left\vert \nu_{1}\right\vert }}{\alpha _{1}}e^{-\alpha_{1}x}%
\right) \right. \right. \right.  \notag \\
&  \notag \\
& \left. \left. \left. +\left( \bar{\mu}\alpha_{2}y+\frac{\sqrt {\left\vert
\nu_{2}\right\vert }}{\alpha_{2}}e^{-\alpha_{2}y}\right) \right] \right\} 
\mathcal{L}_{m}^{2\mu}\left( \frac{2\sqrt{\left\vert \nu _{1}\right\vert }}{%
\alpha_{1}}e^{-\alpha_{1}x}\right) \mathcal{L}_{n}^{2\bar{\mu}}\left( \frac{2%
\sqrt{\left\vert \nu_{2}\right\vert }}{\alpha_{2}}e^{-\alpha_{2}y}\right)
.\right.
\end{align}

We know that $\varepsilon_{mn}=2\xi_{mn}/\hslash^{2}$ and $%
\xi_{mn}=-A+M_{0}(E_{mn}-R)$. Consequently the energy eigenvalues will rise
as solutions of the following transcendental equation%
\begin{align}
& \left. 8~\gamma_{2}\left( \in_{mn}\right) ~\gamma_{4}\left( \in
_{mn}\right) \left( A+\in_{mn}\right) =\gamma_{4}\left( \in_{mn}\right) 
\left[ \left\vert \gamma_{3}\left( \in_{mn}\right) \right\vert -\bar {\alpha}%
_{1}\sqrt{\frac{\gamma_{2}\left( \in_{mn}\right) }{2}}(2n+1)\right]
^{2}\right.  \notag \\
&  \notag \\
& \left. +\gamma_{2}\left( \in_{mn}\right) \left[ \left\vert \gamma
_{3}\left( \in_{mn}\right) \right\vert -\bar{\alpha}_{2}\sqrt{\frac {%
\gamma_{4}\left( \in_{mn}\right) }{2}}(2m+1)\right] ^{2},\right.
\end{align}

\noindent where we defined $\in _{mn}~\equiv M_{0}\left( R-E_{nm}\right) $, $%
\bar{\alpha}_{1}\equiv \hbar ~\alpha _{1}$ and $\bar{\alpha}_{2}\equiv \hbar
~\alpha _{2}$. Considering a symmetrical (in $x$ and $y$) case of the mass
and potential dependencies, in order to have a concrete example to study, we
choose the parameters as given by: $R=0$, $A=M_{0}=g_{1}=g_{3}=\alpha
_{1}=\alpha _{2}=\hslash =1$, $g_{2}=g_{4}=0$, $B_{1}=B_{3}=-1$, $%
B_{2}=B_{4}=1/8$. In this case, the allowed energy levels are given in the
Table below (note that due to the symmetry of the system, the pair $\left(
n,m\right) $ have the same energy as the one $\left( m,n\right) $).
Furthermore, the potential profile appears in the Figure 1 and a plot where
the energy spectrum is presented in scale appears in the Figure 2. Note
that, since the energy of the bound state can not be lower than que smallest
value of the potential and that this potential becomes asymptotically
constant, for the case of the above parameters, the allowed values of the
bound states shall be in the interval $-0.40693\leq E_{n,m}\leq 1.$%
\vspace{0.2cm} 
\begin{table}[h]
\begin{center}
\begin{tabular}{||l||l||l||l||l||l||l||l||l||}
\hline\hline
$n$ & $m$ & $E_{nm}$ & $n$ & $m$ & $E_{nm}$ & $n$ & $m$ & $E_{nm}$ \\ 
\hline\hline
0 & 0 & -0.0669873 & 1 & 3 & -0.161438 & 3 & 4 & 0.250000 \\ \hline\hline
0 & 1 & 0.250000 & 1 & 4 & 0.250000 & 3 & 5 & 0.531754 \\ \hline\hline
0 & 2 & 0.433013 & 2 & 2 & -0.329156 & 3 & 6 & 0.883975 \\ \hline\hline
0 & 3 & 0.250000 & 2 & 3 & -0.116025 & 4 & 4 & 0.410275 \\ \hline\hline
0 & 4 & -0.0188424 & 2 & 4 & 0.170844 & 4 & 5 & 0.631966 \\ \hline\hline
1 & 1 & 0.661438 & 2 & 5 & 0.542893 & 4 & 6 & 0.910275 \\ \hline\hline
1 & 2 & 0.957107 & 3 & 3 & 0.0317542 & 5 & 5 & 0.801042 \\ \hline\hline
\end{tabular}%
\end{center}
\par
\renewcommand{\tablename}{Table}
\caption{Energy levels.}
\end{table}
\vspace{0.1cm} 
\begin{figure}[h]
\includegraphics[scale=0.45]{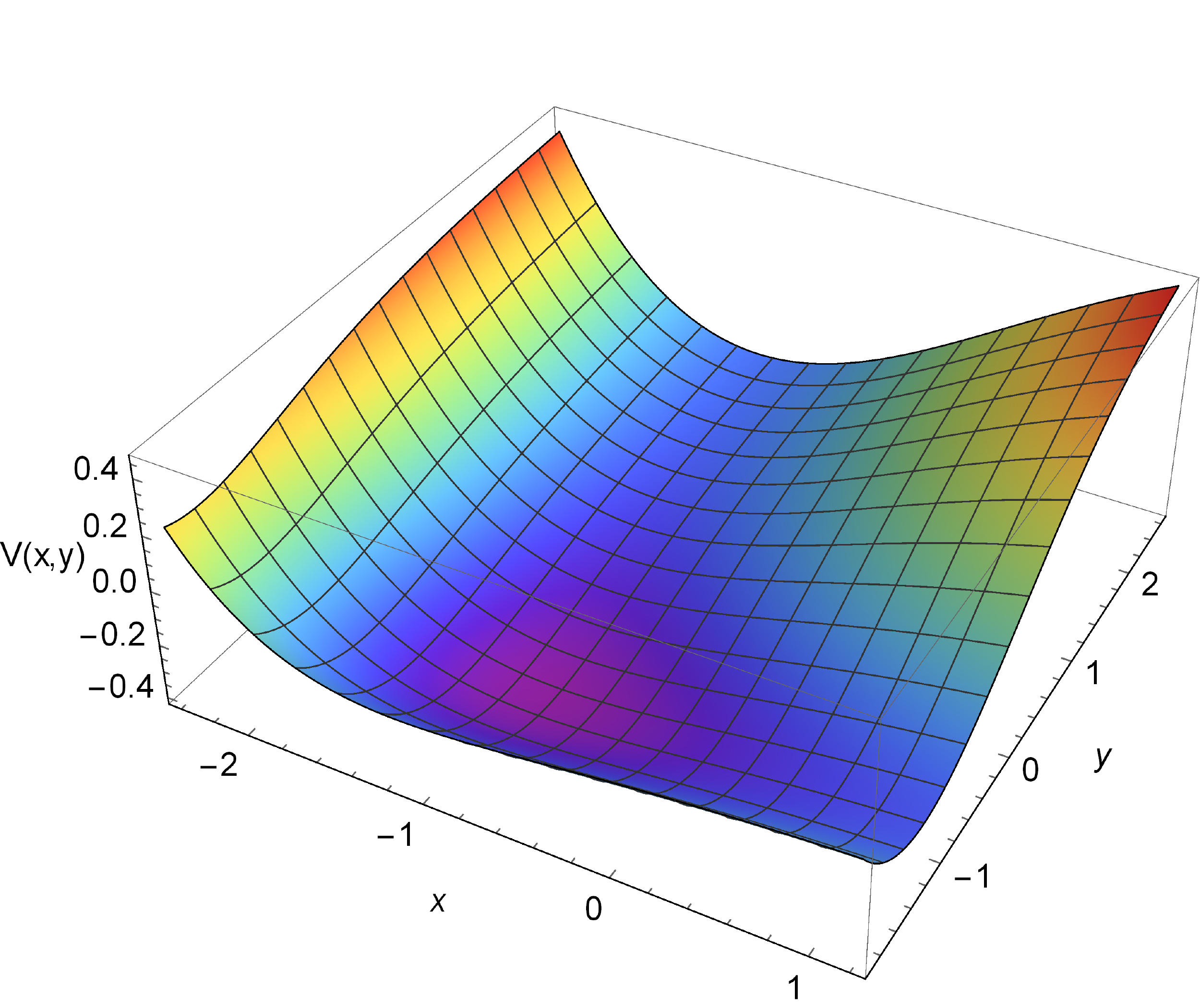}
\caption{Morse-like potential in two dimensions.}
\end{figure}
\begin{figure}[h]
\includegraphics[scale=0.45]{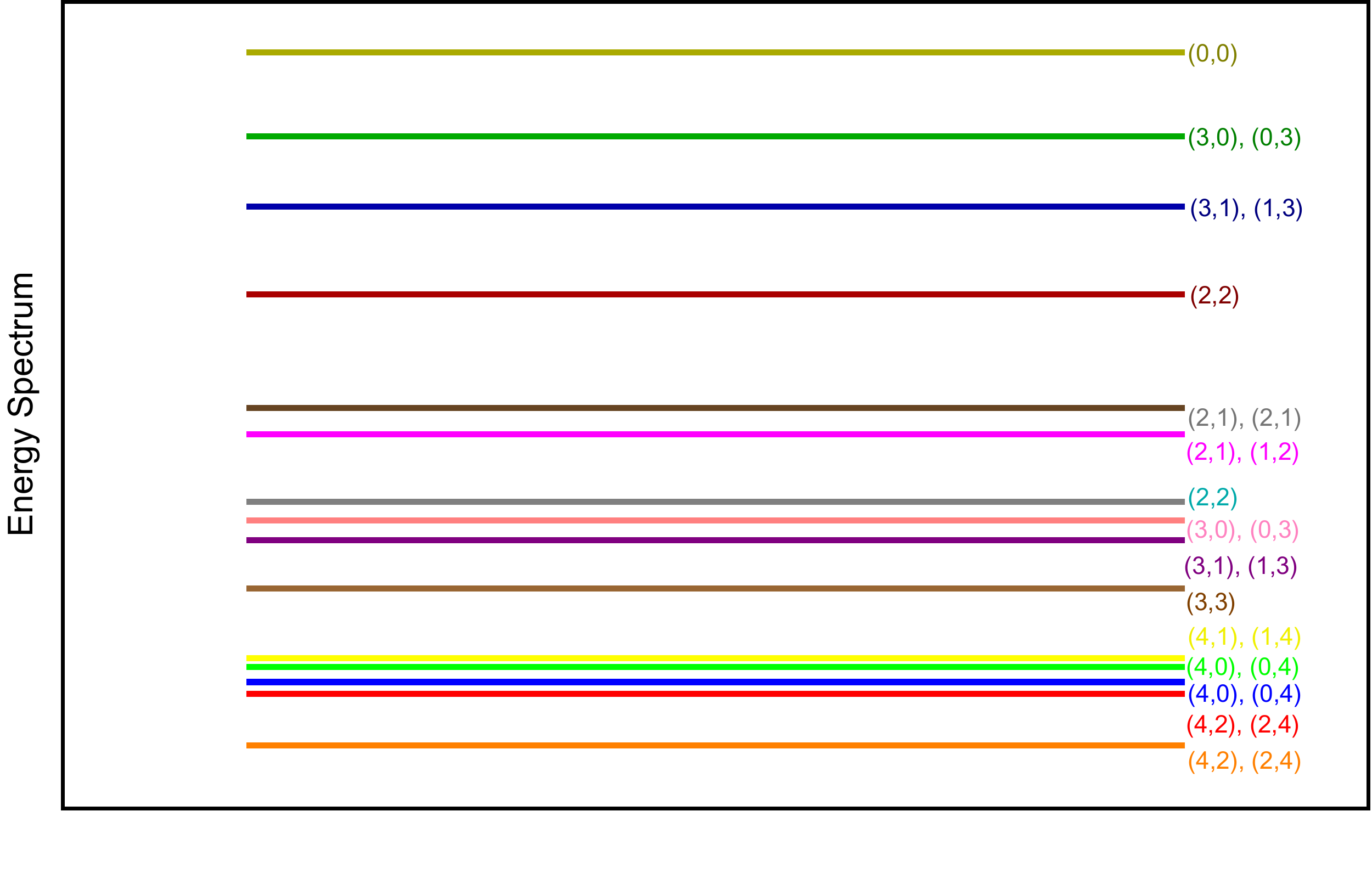}
\caption{Energy spectrum.}
\end{figure}
By observing both the Table I and the Figure 2, one can note that there are
some interesting results in the spectrum. First of all, we observe that this
potential presents a finite number of allowed bound states, which is not a
surprise, since this already happens in the case of the one-dimensional
Morse potential (even in the case with position-dependent masses). However,
in the case analyzed, there are inversions of energies where sates labeled
with higher quantum numbers present lower energies than states with lower
quantum numbers, as happens in the case of atoms with somewhat great atomic
numbers. On the other hand, beyond the some expected degeneracies, we
observe that there is a eight-fold degenerated stated (the seventh exited
one). In this case we checked that one shall have an accidental degeneracy,
since we checked that changing slightly some potential parameters this
degeneracy disappears, becoming a quasi-degeneracy. 

\section{Final comments}

In this work, we present a general construction of a class of a
two-dimensional PDM systems in Cartesian coordinates, analyzing an exactly
solvable case and discussing its ordering ambiguity and some of their
properties. We extend the idea to the problem where ones deal with
increasing mass Morse-like. In this case we obtain the exact wave-functions
and energies for a complete set of eigenstates. Since the energy of the
bound states come from a transcendental equation, involving the quantum
numbers of a pair of one-dimensional equations, we discovered that this
system presents a behavior which emulates the inversion of excited states
usually seen in atoms with high atomic numbers. Moreover, an interesting
accidental degeneracy appeared. Finally, it is important to remark that one
could use this exactly solvable system in order to construct more realistic
ones by using their solution in perturbative approaches. \bigskip

\begin{acknowledgments}
RACC thanks to UNESP-Campus de Guaratinguet\'{a} and CAPES for financial
support. ASD thanks to CNPQ for partial financial support, and J.A.O. thanks
to DFQ of UNESP, Campus de Guaratinguet\'{a}, where this work was carried
out.
\end{acknowledgments}


\noindent

\begin{thebibliography}{99}
\bibitem{Alhassid} Y. Alhassid, Rev. Mod. Phys. \textbf{72 }(2000) 895.

\bibitem{rmp1} M. A. Kastner, Rev. Mod. Phys. \textbf{64} (1992) 849.

\bibitem{prl0} A. R. Wright, M. Veldhorst, Phys. Rev. Lett. \textbf{111}
(2013) 9, 096801.

\bibitem{prb1} G. Anatoly, Phys. Rev. B \textbf{91} (2015) 20, 205105.

\bibitem{prb2} Y. Li, A. Kundu, F. Zhong, and B. Seradjeh, Phys. Rev. 
\textbf{90} (2014) 12, 121401.

\bibitem{pre1} B. Wahlstrand, I. I. Yakimenko, K.-F. Berggren, Phys. Rev. E 
\textbf{89} (2014) 6, 062910.

\bibitem{prl1} P. Tighineanu, M. L. Andersen, A. S. Sorensen, S. Stobbe, and
P. Lodahl, Phys. Rev. Lett. \textbf{113} (2014) 043601.

\bibitem{Lozada-Dong-Yu} M. Lozada-Cassou, S. H. Dong, and J. Yu, Phys.
Lett. A \textbf{331} (2004) 45.

\bibitem{Schmidt-Azeredo-Gusso} A. G. M. Schmidt, A. D. Azeredo, and A.
Gusso, Phys. Lett. A \textbf{372} (2008) 2774.

\bibitem{BenDaniel-Duke} D. J. BenDaniel and C. B. Duke, Phys. Rev. B 
\textbf{152} (1966) 683.

\bibitem{Gora-Williams} T. Gora and F. Williams, Phys. Rev. \textbf{177}
(1969) 1179.

\bibitem{Bastard} G. Bastard. Phys. Rev. B \textbf{24} (1981) 5693.

\bibitem{Ross} O. Von Roos, Phys. Rev. B \textbf{27} (1983) 7547.

\bibitem{Zhu-Kroemer} Q. G. Zhu and H. Kroemer, Phys. Rev. B \textbf{27}
(1983) 3519.

\bibitem{Li-Kuhn} T. L. Li and K. J. Kuhn, Phys. Rev. B \textbf{47} (1993)
12760.

\bibitem{Cavalcante-Filho-Almeida-Freire} F. S. A. Cavalcante, R. N. Costa
Filho, J. Ribeiro Filho, C. A. S. de Almeida, and V. N. Freire, Phys. Rev. B 
\textbf{55} (1997) 1326.

\bibitem{Dutra-Oliveira} A. de Souza Dutra and J. A. de Oliveira, J. Phys.
A: Math Theor. \textbf{42} (2009) 025304.

\bibitem{Shewell} J. R. Shewell, Am. J. Phys. \textbf{27} (1959) 16.

\bibitem{Slater} J. C. Slater, Phys. Rev. \textbf{76} (1949) 1592.

\bibitem{Luttinger} J. M. Luttinger and W. Kohn, Phys. Rev. \textbf{97}
(1955) 869.

\bibitem{Wannier} G. H. Wannier, Phys. Rev. \textbf{52} (1957) 191.

\bibitem{rojo} \'{O}. Rojo and J. S. Levinger, Phys. Rev. \textbf{123}
(1961) 2177.

\bibitem{razavy} M. Razavy, G. Field, and J. S. Levinger, Phys. Rev. \textbf{%
125} (1962) 269.

\bibitem{bastard} G. Bastard, Wave Mechanics Applied to Semiconductor
Heterostructres, Les \'{E}ditions de Physique, Les Ullis, 1992.

\bibitem{weisbuch} C. Weisbuch and B. Vinter, Quantum Semiconductor
Heterostructures, Academic Press, New York, 1993.

\bibitem{ref1} C. C. Wu, J. Sun, F. J. Huang, Y. D. Li, W. M. Liu, Europhys.
Lett. \textbf{104} (2013) 27004.

\bibitem{ref2} M. Benito, A. G\'{o}mez-Le\'{o}n, V .M. Bastidas, T. Brandes,
and G. Platero, Phys. Rev. B \textbf{90} (2014) 20, 205127.

\bibitem{ref3} J. Gong and Qing-hai Wang, Phys. Rev. A \textbf{91} (2015) 4,
042135.

\bibitem{pla2000} A. de Souza Dutra and C. A. S. de Almeida, Phys. Lett. A 
\textbf{275} (2000) 25.

\bibitem{m1} C. Quesne and V. M. Tkachuk, J. Phys. A \textbf{37} (2004) 426.

\bibitem{m2} J. F. Carinena, M. F. Ranada, and M. Santander, Ann. Phys. 
\textbf{322} (2007) 434.

\bibitem{m3} A. Ganguly and L. M. Nieto, J. Phys. A \textbf{40} (2007) 7265.

\bibitem{m4} S. Choi, K. M. Galdamez, and B. Sundaram, Phys. Lett. A \textbf{%
374} (2010) 3280.

\bibitem{m5} A. Arda, R. Sever, and C. Tezcan, Phys. \ Scripta \textbf{79}
(2009) 015006.

\bibitem{m6} A. Ganguly and A. Das, J. Math. Phys. \textbf{55} (2014) 11,
112102.

\bibitem{jpa2006} A. de Souza Dutra, J. Phys. A \textbf{39} (2006) 203.

\bibitem{dutra2} A. de Souza Dutra, M. B. Hott, and C. A. S. Almeida,
Europhys. Lett. \textbf{62} (2003) 8.

\bibitem{schimidt1} A. G. M. Schmidt, Phys. Lett. A \textbf{353} (2006) 459.

\bibitem{quesne} C. Quesne, J. Math. Phys. \textbf{49} (2008) 022106; J.
Phys. A \textbf{40} (2007) 13107; Ann. Phys. \textbf{321} (2006) 1221.

\bibitem{sever} S. M. Ikhdair and R. Sever, J. Mol. Str. Theochem \textbf{885%
} (2008) 13.

\bibitem{halberg} A. Schulze-Halberg, Int. J. Mod. Phys. A \textbf{22}
(2007) 1735; \textbf{21} (2006) 4853; \textbf{21} (2006) 1359.

\bibitem{carinena} J. F. Carinena, M. F. Ranada and M. Santander, Ann. Phys. 
\textbf{322} (2007) 2249.

\bibitem{ganguly} A. Ganguly and L. M. Nieto, J. Phys. A \textbf{40} (2007)
7265.

\bibitem{roy} B. Roy, Mod. Phys. Lett. B \textbf{20} (2006) 1033.

\bibitem{mustafa} O. Mustafa and S. H. Mazharimousavi, J. Phys. A \textbf{41}
(2008) 244020; \textbf{39} (2006) 10537.

\bibitem{tanaka} T. Tanaka, J. Phys. A \textbf{39} (2006) 219.

\bibitem{koca} R. Koc, M. Koca, and G. Shaninoglu, Eur. Phys. J. B \textbf{48%
} (2005) 583.

\bibitem{znojil} U. Gunther, F. Stefani, and M. Znojil, J. Math. Phys. 
\textbf{46} (2005) 063504.

\bibitem{cg1} G. Chen, Chin. Phys. \textbf{14} (2005) 460.

\bibitem{Cg2} G. Chen and Z. D. Chen, Phys. Lett. A \textbf{331} (2004) 312.%
{\small \ }

\bibitem{a0} A. A. Stahlhofen, J. Phys. A \textbf{37} (2004) 10129-10138,

\bibitem{a1} B. Bagchi, P. Gorain, C. Quesne, and R. Roychoudhury, Mod.
Phys. Lett. A \textbf{19} (2004) 2765-2775,

\bibitem{a3} K. Bencheikh, K. Berkane, and S. Bouizane, J. Phys. A \textbf{37%
} (2004) 10719.

\bibitem{a4} J. A. Yu and S. H. Dong, Phys. Lett. A \textbf{325} (2004) 194.

\bibitem{a5} C. Quesne and V. M. Tkachuk, J. Phys. A \textbf{37} (2004) 4267.

\bibitem{a6} Y. C. Ou, Z. Q. Cao, and Q. H. Shen, J. Phys. A \textbf{37}
(2004) 4283.

\bibitem{a7} R. Koc and H. Tutunculer, Annalen der Physik \textbf{12} (2003)
684.

\bibitem{a8} A. D. Alhaidari, Phys. Rev. A \textbf{66} (2002) 042116.

\bibitem{a9} B. Roy and P. Roy, J. Phys. A \textbf{35} (2002) 3961.

\bibitem{npb} S. Ramgoolam, B. Spence, and S. Thomas, Nucl. Phys. B \textbf{%
703} (2005) 236.

\bibitem{epl05} A. de Souza Dutra, M. B. Hott, and V. G. C. S. dos Santos,
Europhys. Lett. \textbf{71} (2005) 166.

\bibitem{pla87} B. K. Cheng and A. de Souza Dutra, Phys. Lett. A \textbf{123}
(1987) 105.

\bibitem{pra89} A. de Souza Dutra and B. K. Cheng, Phys. Rev. A \textbf{39}
(1989) 5897.

\bibitem{pla91} A. de Souza Dutra, C. F. de Souza, and L. C. de Albuquerque,
Phys. Lett. A \textbf{156} (1991) 371.

\bibitem{abdalla2007} M. S. Abdalla and J. R. Choi, Ann. Phys. \textbf{322}
(2007) 2795.

\bibitem{dutra-juliano} A. de Souza Dutra and J. A. de Oliveira, J. Phys. A:
Math. Theor. \textbf{42} (2009) 025304.

\bibitem{cc1} A. G. M. Schmidt, J. Phys. A: Math. Theor. \textbf{42} (2009)
245304.

\bibitem{cc2} O. Mustafa, J. Phys. A: Math. Theor. \textbf{43} (2010) 385310.

\bibitem{cc3} A. G. M. Schmidt, Phys. A \textbf{391}(2012) 3792.

\bibitem{cc4} A. G. M. Schmidt, L. Portugal, and A. L. de Jesus, J. Math.
Phys. \textbf{56} (2015) 012107.
\end{thebibliography}
\end{document}